# Network Centrality Analysis of Tehran Urban and Suburban Railway System

**Mohieddin Jafari[1] and Sayed Mohammad Fakhar[2]**
1-Assistant Professor, Institute for Research in Fundamental Sciences (IPM) and Pasteur Institute of Iran (IPI).
2-Independent Researcher, and Journalist at HAMSHARI newspaper

**Abstract**
Nowadays, Tehran Urban and Suburban Railway System (TUSRS) is going to be completed by eight lines and 149 stations. This complex transportation system contains 168 links between each station pairs and 20 cross-section and Y-branch stations among all eight lines. In this study, we considered TUSRS as a complex network and undertook several analyzes based on graph theory. Examining e.g. centrality measures, we identified central stations within TUSRS. This analysis could be useful for redistributing strategy of the overcrowded stations and improving the organization of maintaining system. These findings are also promising for better designing the systems of tomorrow in other metropolitan areas in Iran.

[1]Assistant professor, 09122991069, mjafari@ipm.ir
[2]Journalist, 09128401078, mohamadfakhar@gmail.com

# 1-Introduction

A transportation problem is regarded as a first published paper in the field of graph theory in 1736. The name of this problem was Seven Bridges of Königsberg which was proved unsolvable by Leonhard Euler (1). However, in the development of graph theory and more generally network science, the transportation issue is not well contributed as well as social or biological questions. Nowadays, the transit system are growing and getting bigger, more complex and intricate which study of them is required using graph theory and network analysis approach. By exploring network features, it could be possible to help planners so as to propose better and more robust design of the transit systems of tomorrow (2-5).

Among the network analysis methods, centrality analysis could explain well about one fundamental aspect of a transit system (2). Which station is the most important one in the system? There are several centrality measures which describe central, influential or important nodes in different network structure (6, 7).

# 2-Research Questions/Problems and Objectives

Tehran Urban and Suburban Railway System (TUSRS) is a rapid transit system serving of the capital of Iran. This system consists of eight operational metro lines with construction under way on two lines including line 6 and line 7. On average, more than three millions passengers/day use the TUSRS and it would be increased by completing remaining lines. TUSRS is one of the largest and most crowded subway system in the world which could be considered as a large complex network (8, 9).

In this study, we reconstructed the network model of TUSRS and tried to find central nodes (i.e. stations) within TUSRS network. The central stations are inferred based on several well-known measures. Finally, we compared the potential of all lines and some major stations individually.

## 3-Results

TUSRS network is really a complex network which is interested to study concisely in the aspect of network science. The network view of TUSRS has schemed in Fig. 1. In this figure, node size is proportional to degree centrality measures whereas cross-section stations (no. 16) such as Teatr-e ShahrndEmamKomeini have degree of four and the biggest size, Y-branch stations (no. 4) such as Eram-e Sabz and Shahed have degree of three and the big size, initial stations (no. 15) such as Tajrish and Ghaem have degree of one and small size and the others with the medium size and degree of two. The stations (nodes) are sorted clockwise in two circles according to closeness centrality measure. Top 24 stations based on closeness centrality measure are sorted at the inner circle. Also, a color spectrum from green to brown was used to display the value of betweenness centrality measure in TUSRS.

This three measure is also demonstrated in a matrix pairwise plot (Fig. 2). The first row of this matrix plot indicates the distribution of four centrality measures in all eight lines with different colors. As is shown, the line 4 and 3 of TUSRS demonstrate the highest values according to closeness and betweenness and lowest values in eccentricity measures. This indicates the importance of these two lines among the whole system of TUSRS. In the first column of the matrix plot, this issue is repeated using box plot and five-number summary of each measure. The co-association of the measures also plotted in Fig. 2 to show the independent behavior of the measures. Except for the negative relationship between closeness and eccentricity measures, the independent patterns were observed among the others. Note that the diameter of this figure just shows the range of each attribute.

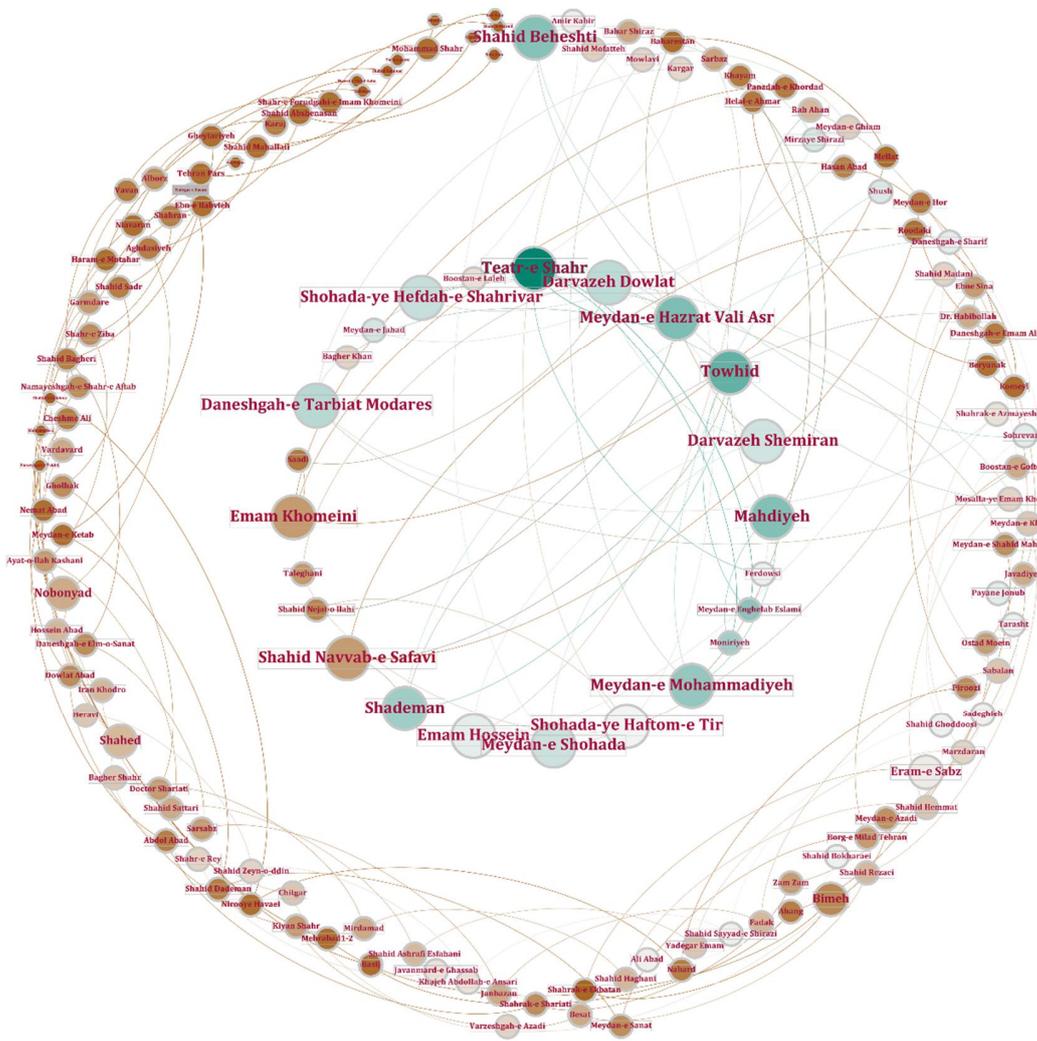

*Figure 1: An overall picture of TUSRS network. In this figure, stations (nodes) are sorted clockwise in two circles according to closeness centrality measure. Also, node size and color are proportional to betweenness and degree centrality measures respectively.*

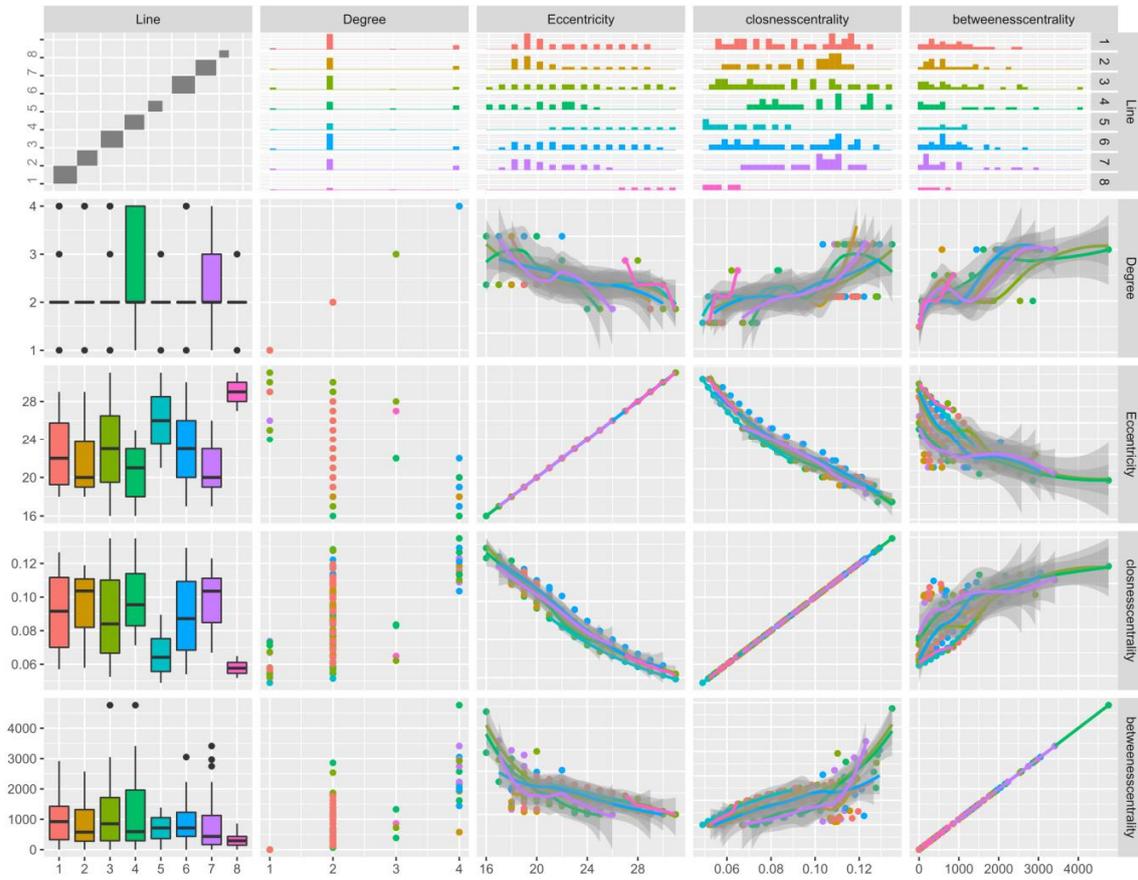

*Figure 2: A matrix plot of four centrality measures computed in all eight lines of TUSRS.*

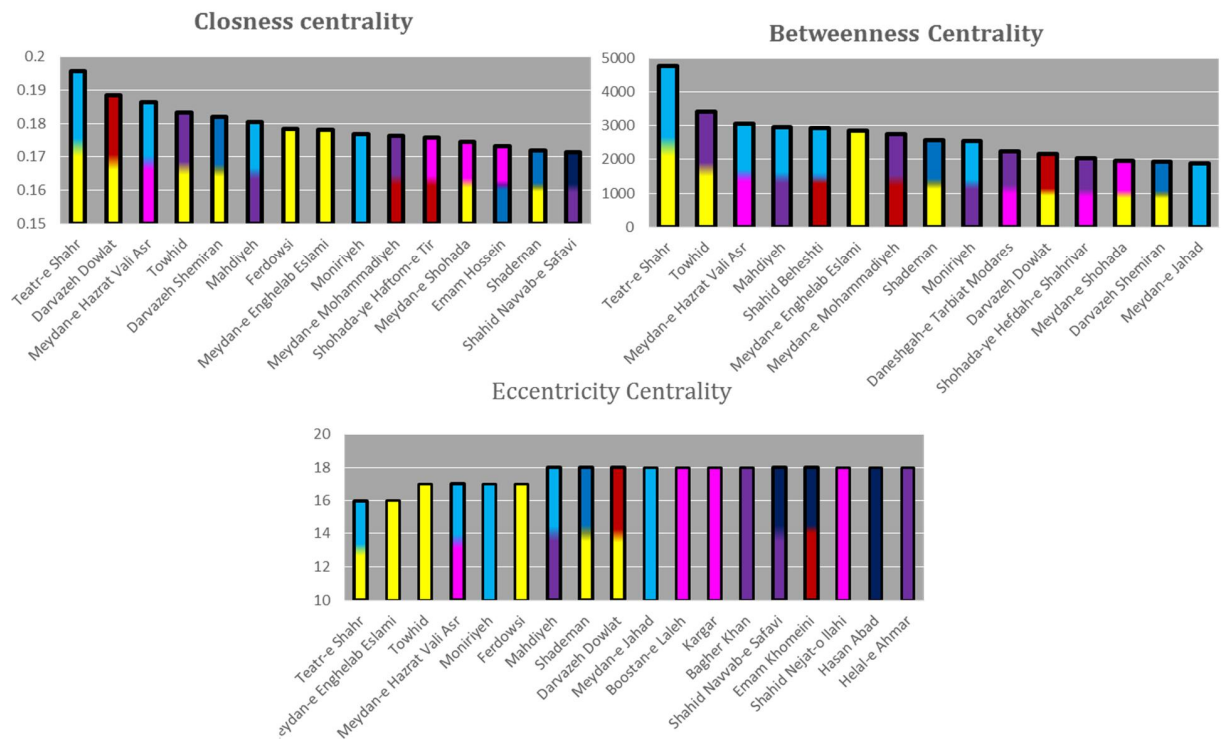

*Figure 3: Most central stations according to closeness, betweenness and eccentricity centrality measures. The colors indicate the line number (1: red, 2: dark blue, 3: blue, 4: yellow, 5: dark green, 6: pink, 7: purple, 8: green ).*

To explain more concisely, top stations based on closeness, betweenness and eccentricity are displayed in Fig. 3. The bar plot presented in these figure facets are colored based on the line color in TUSRS. Same as two other measures, Teatr-e Shahr is the top rank in closeness centrality measure. The station that has the highest value of this criterion is a station that is close to all stations and if start from each point, this station is the closest one on average. Interestingly, the oldest cross-section station i.e. Emam Khomeini is not presented among the top ten percent of the closeness centrality measure and also three stations with the degree of two that they are not cross-section present in the top list i.e. Ferdowsi, Meydan-e EnghelabEslami and Moniriyeh. This ranking is very useful for routing algorithm to propose the shortest path from any origin to destination stations. This is also applicable for locating shopping center near to all metro ridership or defining a more accessible place for any purpose.

The second facet of the Fig. 3 shows the ranking of the stations according to betweenness centrality measure which indicates the potential of each station to monitor the relationship between other stations. This measure draws the particular attention to infrastructure maintenance and sustainability of these stations, including elevators, escalators and public facilities. Interestingly, non-cross-section stations i.e. Meydan-e EnghelabEslami and Meydan-e Jahad also exist among the top rank. Based on this measure, it could be concluded that monitoring and security controlling in these stations are more essential. The betweenness centrality measure could also be computed for the edges (station links) within a network. Table 1 shows the top-ranked list of the links which remarkably include Meydan-e EnghelabEslami station in the first and second ranks.

The last measure is eccentricity which represents the station reaching potential to others. In this measure, the lowest value represents the average number of stations between the given station and any other stations. Again the stations of the line 4 are more significant. This measure is also useful for any public services e.g. rescue unit to reach other station using TUSRS railways.

*Table 1: The edge betweenness centrality measures of TUSRS.*

| From | To | Value |
|---|---|---|
| Teatr-e Shahr | Meydan-e EnghelabEslami | 2960.5 |
| Towhid | Meydan-e EnghelabEslami | 2906.5 |
| Teatr-e Shahr | Moniriyeh | 2639.833333 |
| Mahdiyeh | Moniriyeh | 2582.5 |
| Meydan-e Hazrat Vali Asr | Teatr-e Shahr | 2461.5 |
| Towhid | Shademan | 2425.5 |
| Mahdiyeh | Meydan-e Mohammadiyeh | 2162.666667 |
| Meydan-e Hazrat Vali Asr | Meydan-e Jahad | 1995.666667 |
| MirzayeShirazi | Meydan-e Jahad | 1897.666667 |
| DarvazehShemiran | DarvazehDowlat | 1890 |
| Meydan-e Mohammadiyeh | Shush | 1890 |

In the present map of TUSRS, the cross-section between line 2 and 3 in Emam Ali station was omitted. It had not happened otherwise this station would have ranked second and third stations in closeness and betweenness centrality measure top ranks (data not shown). This issue could be important regarding the line 5 and line 2 cross-section which is the entrance of Karaj suburban ridership. Based on this alteration, average centrality measures of line 2 decreased, instead line 4 become comparatively more important and central in the TUSRS especially regarding the betweenness centrality measure.

In conclusion, it should be rephrased that graph theory-based study of the transient system helps to the better understanding of the system and fruitful for planners and related organization. Note that integrating similar urban transient systems such as Bus Rapid Transient (BRT) with TUSRS could be more promising for modern designing the human mobility network within the cities. Regarding the growing urban population, this approach is applicable in accordance with the sustainable development of cities.

## 4-References